# The WebShop E-Commerce Framework

**Marcus Fontoura**
IBM Almaden Research Center
650 Harry Road, San Jose, CA 95120, U.S.A.
e-mail: fontoura@almaden.ibm.com

**Wolfgang Pree**
Professor of Computer Science
Software Research Lab
University of Constance, D-78457 Constance, Germany
e-mail: pree@acm.org

**Bernhard Rumpe**
Software and Systems Engineering
Munich University of Technology, D-80290 Munich, Germany
e-mail: rumpe@acm.org

*Abstract - This paper presents an e-commerce framework called WebShop, which was developed by the authors for the purpose of demonstrating the use of UML and the UML-F in the domain of Web applications. Thus, the WebShop is not regarded as a full-fledged system out of which real Web stores can be derived. For example, the framework in the presented version does not encounter security features. However, it presents the most important variation points related to online catalogs. The UML-F Web site http://www.UML-F.net provides the Java source files and some sample adaptations of WebShop.*

## 1. WebShop overview

WebShop basically allows the creation of online stores from a description of the products that should be offered and sold on a Web site (see Figure 1).

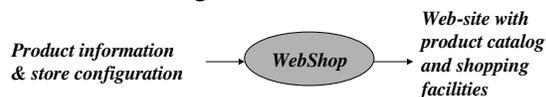

Figure 1 The goal of the WebShop framework.

As specific Web stores differ in various aspects, the WebShop framework defines the following variation points:

- Payment options: companies accept various payment options, such as credit cards, electronic money, and so on. Moreover, completely new electronic payment methods may arise and the framework should be able to incorporate them.
- Promotions: promotions usually depend on parameters such as the overall shopping volume of a customer or the frequency a customer comes along. For example, a bookstore site might send a gift at the end of the year if the sales volume of a customer has surpassed a certain limit. Another example would be to freely upgrade to a faster delivery, if a customer buys goods for a greater value amount. WebShop should be easily extended in that regard.
- Reports: every organization requires different kinds of management information. Examples include rankings of the best customers, sales figures on various single products and product groups, and information regarding the preferred payment methods. Once again, WebShop should be open for any extension of the reporting subsystem.

Figure 2 sketches the configuration options of WebShop. Typical WebShop adaptations can choose among predefined payment options so that this aspect allows a black-box configuration. The promotion and report generation will quite likely be adapted to the specific requirements of each application.

Figure 3 represents the navigational structure of a typical WebShop application. Each single rectangle represents a web page and the arrows represent the actions that cause movement between the pages.

The Shopping page is the application entry point. It displays the list of available products,

allowing clients to add products to their individual shopping cart and to change the quantities of each selected product. When the client wants to checkout he or she only has to select payment method and to provide the required payment information. The system then verifies the information and either processes the transaction or reports an error. Figure 4 shows a typical Shopping page.

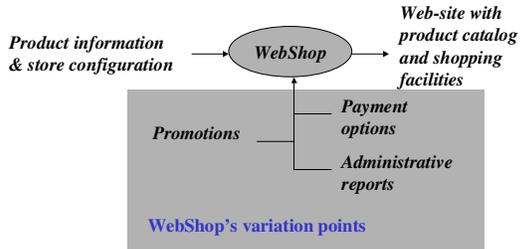

Figure 2. Variation points of WebShop.

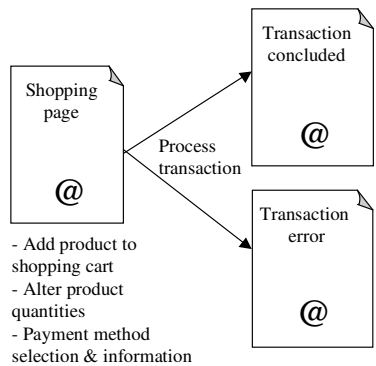

Figure 3. Navigational structure of a typical WebShop application.

WebShop allows the creation of a complementary site for displaying the administrative reports. Typically, the structure of the administrative is set up as a simple list of reports. The end user can select from an overview list any of the reports available (see Figure 5).

## 2. WebShop components

The following sections present the core aspects of WebShop by means of UML-F [1] diagrams. UML-F is an UML [4] extension for documenting object-oriented frameworks and design patterns. This presentation forms the basis for identifying patterns that are useful in the context of e-commerce frameworks.

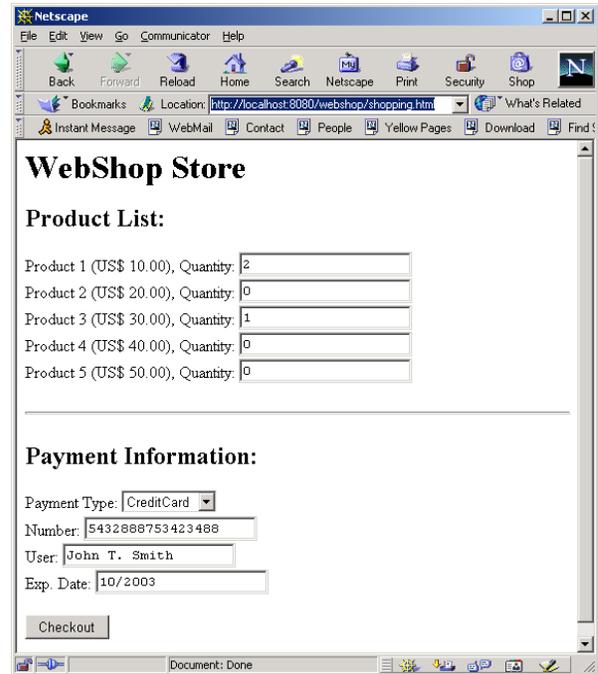

Figure 4. Typical Shopping page for WebShop.

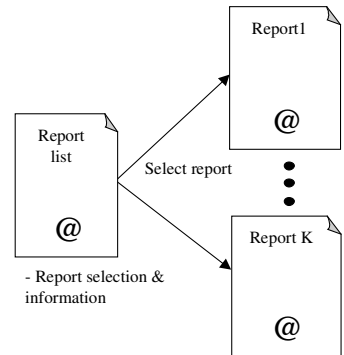

Figure 5. Report listing on a separate site.

### 2.1. Shopping Cart

The core entity of the WebShop framework is the shopping cart. For each client access to a Web store a new shopping cart object is created. This shopping cart takes care of the connection and is responsible for controlling the user selection of products and the checkout operation. Figure 6 shows an UML-F diagram demonstrating what a shopping cart contains - the exact number of products and one transaction log.

Methods addProduct(), removeProduct(), and changeQuantity() in class ShoppingCart modify the products already chosen accordingly. Method checkout() processes the payment transaction.

It is also responsible for updating the system transaction log by invoking method addLog() in class TransLog. The transaction log may be used for various customer relationship management activities such as promotions. Thus, it forms the basis of various reports.

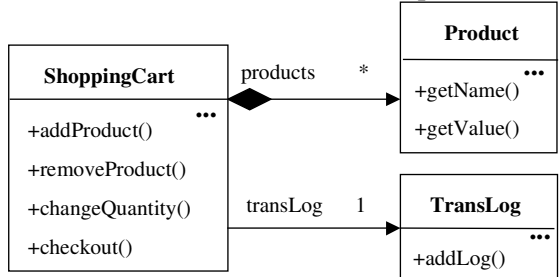

Figure 6. UML-F class diagram of ShoppingCart and two of its associated classes.

### 2.2. Payment options

Each application created by the framework will incorporate a number of payment options. In particular, the cart's checkout() method requires the information on available payment choices. To keep payment methods flexible, WebShop applies the Separation construction principle [3] (see Figure 7). As all the Payment objects interacting with a shopping cart have to be able to process a payment, the interface Payment defines the processPayment() method. Therefore, all specific classes used for payment have to implement the Payment interface, as illustrated in Figure 7.

Figure 7 uses UML-F tags to identify explicitly the template and hook methods and classes. The method checkout() is the template method («Sep-t»), since it is responsible for invoking processPayment() («Sep-h»), which is a hook method that varies for different classes that implement the Payment interface. The «adapt-dyn» tag indicates that the classes are dynamically loaded into the system when needed.

From the client's perspective, a Web form should present the payment options available. The client selection is then proceeded to the store that has to instantiate the appropriate payment object and plug it into the shopping cart. The sequence diagram in Figure 8 illustrates this scenario. It shows the creation of an object to process credit card transactions in Figure 8(a), and the electronic money transactions in Figure 8(b).

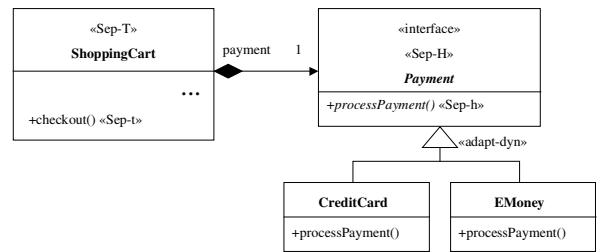

Figure 7. The Payment interface.

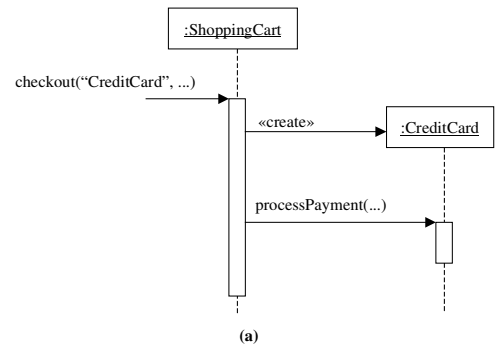

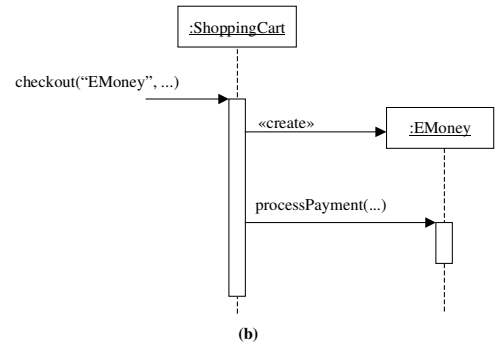

Figure 8. Creating the appropriate payment object (a) for credit cards and (b) for electronic money.

Note that the parameter specifying the payment option is a string that represents the class name of the specific payment class. The checkout() method uses dynamic class loading to instantiate the appropriate class based on its name. Example C.1 illustrates the code for the checkout(). A more elaborate (and flexible) design would have a table mapping the string parameters to the actual class names.

```
public boolean checkout(
  String paymentClassName,
  String paymentInfo) {

  boolean paymentOK = false;
```

```
  Payment payment = null;

  try {  // tries to instantiate a Payment object
    Class c = Class.forName(paymentClassName);
    payment = (Payment) c.newInstance();
  }
  catch(Exception e) {
    // error, throws framework exception
  }
   // The method total() calculates the total value
   // of goods in the shopping cart. This method is a
   //private method in class ShoppingCart.

   if (payment.verifyPayment(payInfo, total())) {
     // Add transaction to log
     paymentOK = true;
   }
   return paymentOK;
}
```

Example 1.  Source code fragments of method checkout() in class ShoppingCart.

The various implementations of the processPayment() hook method require different arguments. As the client supplies these arguments through a Web form, WebShop assumes that they are provided in a single string that is formatted according to simple conventions, that is, as "number = '5534453567144532'; expdate = '10/2002'; name = 'John V. Lee'". (Another equivalent solution would is to use XML for formatting this input string). Each implementation of processPayment() parses and processes this input string. The attributes in the string are defined by the particular payment classes. The sequence diagram in Figure 9 illustrates this behavior for the CreditCard object.

### 2.3. Defining promotions in WebShop

In order to deal with promotions, the checkout() method invokes a method definePromo(). This method defines a promotion that depends, for example, on the overall value of purchased goods.

As the WebShop framework should be able to support several promotions at the same time, WebShop applies the Chain-of-Responsibility (COR) pattern [2]. For example, when a frequent shopper buys goods for more than $ 1000.00 he or she should receive an extra discount.

The COR pattern allows each promotion object to check if the current transaction follows the conditions required by it. The object then forwards the request to the next promotion object, if any. The object diagram in Figure 10 exemplifies a combination of two such promotion objects to which a ShoppingCart object refers to.

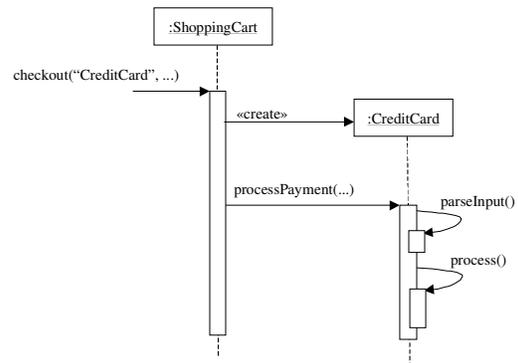

Figure 9. The general behavior of concrete implementations of processPayment().

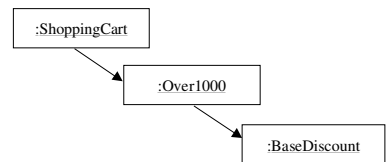

Figure 10.  Composing promotion objects.

The ShoppingCart object is responsible for invoking the definePromo() method as the first in the chain of promotion objects. These promotion objects are further responsible for forwarding the request in the chain. In the sample chain shown in Figure C.10, the object of class Over1000, which gives discounts for transactions over $ 1000.00, treats the request and forwards it to the next promotion object, which gives a 10% discount for all transactions. Figure 11 annotates the promotions variation point with the COR tags.

Of course, the solution based on COR is simple and doesn't take in account some important issues, such as the fact that some implementations of definePromo() should have access to the payment and user information and to the transaction log in order to support promotions based on the user history, such as promotions for frequent shoppers. However, COR provides the main structure of this variation point to which new extensions can be added.

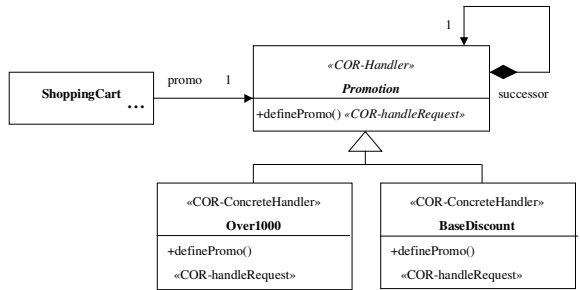

Figure 12. Annotation of the promotions variation point with the COR tags.

## 2.4. Reports

The report generation in WebShop relies on the Separation construction principle (see Figure 13). Analogous to the Payment subclasses, a string is used to uniquely specify which class implementing Report should be loaded the system, as illustrated in Figure 14.

The report() method is responsible for the dynamic loading of the appropriate ReportImp class and for the invocation of the generateReport() method, which returns a string containing the report written/generated in HTML. For example, class ListLog lists the entire log while class ByProduct might lists the transactions related to a given product.

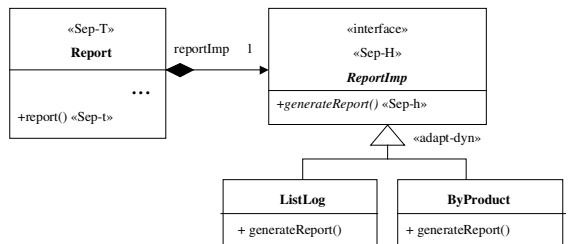

Figure 13. Annotating the report variation point through the Separation construction principle tags

## 2.5. The Web Request pattern

As the design used for the Report variation point is analogous to the one used for Payment, and as it is quite useful to have a dynamically created object to treat end-user requests, we define a domain-specific pattern called Web Request. The payment and administrative reports variation points both use the Web Request structure. Payment, for instance, is a specialization of Web Request that treats payment processing requests. Figure 15 models the administrative reports variation point together with the Web Request tags (the same can be done to payment).

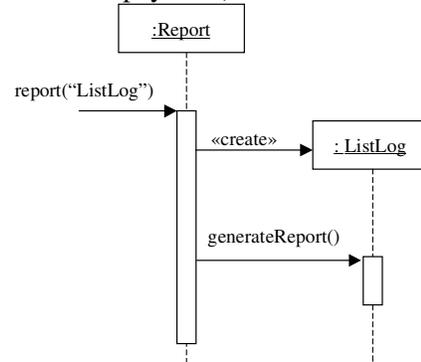

Figure 14. Creating the appropriate report object

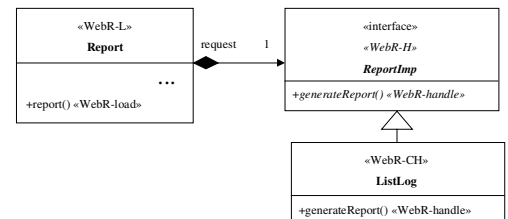

Figure 15. Using the Web Request tags to model the report generation variation point.

Web Request keeps the execution of a request flexible and applies the Separation construction principle [2] for this purpose. The «WebR–load» method is responsible for dynamic loading of the appropriate «WebR–ConcreteHandler» class based on its input arguments

## 3. Conclusions

The paper presented the core components of the WebShop framework, which focus assisting the development of online stores. The UML extensions, namely the UML-F profile [1] used to describe WebShop proved very effective to provide an application developer an intuitive and easy overview of the framework. The UML-F profile mainly provides a set of tags together with mechanisms to introduce new tags and to describe their meaning and intention in an informal yet systematic way.